\documentclass[%
superscriptaddress,
 amsmath,floatfix,amssymb,
 aps,
onecolumn
]{revtex4-1}

\usepackage{graphicx}
\usepackage{amssymb}
\usepackage{float}
\usepackage{amsmath}
\usepackage{verbatim}
\usepackage{graphicx}
\usepackage[toc,page]{appendix}
\usepackage{braket}
\usepackage{amsmath}
\usepackage{siunitx,esvect}
\usepackage{siunitx}
\usepackage{textcomp}
\usepackage{gensymb}
\usepackage{appendix}
\usepackage{braket}
\usepackage{upgreek}
\usepackage[breaklinks=true,colorlinks=true,linkcolor=blue,urlcolor=blue,citecolor=blue]{hyperref}
\usepackage{dcolumn}
\usepackage{bm}
\usepackage{natbib}
\usepackage{multirow}

\setcitestyle{super}

\begin{document}

\title{Entanglement demonstration on board a nano-satellite}
\author{Aitor Villar\footnote{Email address: aitor.villar@u.nus.edu.}}
\affiliation{%
 Centre for Quantum Technologies, National University of Singapore, 3 Science Drive 2, S117543, Singapore\\
}%
\author{Alexander Lohrmann}%
\affiliation{%
 Centre for Quantum Technologies, National University of Singapore, 3 Science Drive 2, S117543, Singapore\\
}%
\author{Xueliang Bai}%
\affiliation{%
 Centre for Quantum Technologies, National University of Singapore, 3 Science Drive 2, S117543, Singapore\\
}%
\author{Tom Vergoossen}%
\affiliation{%
 Centre for Quantum Technologies, National University of Singapore, 3 Science Drive 2, S117543, Singapore\\
}%
\author{Robert Bedington}%
\affiliation{%
 Centre for Quantum Technologies, National University of Singapore, 3 Science Drive 2, S117543, Singapore\\
}%
\author{Chithrabhanu Perumangatt}%
\affiliation{%
 Centre for Quantum Technologies, National University of Singapore, 3 Science Drive 2, S117543, Singapore\\
}%
\author{Huai Ying Lim}%
\affiliation{%
 Centre for Quantum Technologies, National University of Singapore, 3 Science Drive 2, S117543, Singapore\\
}%
\author{Tanvirul Islam}%
\affiliation{%
 Centre for Quantum Technologies, National University of Singapore, 3 Science Drive 2, S117543, Singapore\\
}%
\author{Ayesha Reezwana}%
\affiliation{%
 Centre for Quantum Technologies, National University of Singapore, 3 Science Drive 2, S117543, Singapore\\
}
\author{Zhongkan Tang}%
\affiliation{%
 Centre for Quantum Technologies, National University of Singapore, 3 Science Drive 2, S117543, Singapore\\
}
\author{Rakhitha Chandrasekara}%
\affiliation{%
 Centre for Quantum Technologies, National University of Singapore, 3 Science Drive 2, S117543, Singapore\\
}
\author{Subash Sachidananda}%
\affiliation{%
 Centre for Quantum Technologies, National University of Singapore, 3 Science Drive 2, S117543, Singapore\\
}
\author{Kadir Durak}%
\affiliation{%
 Centre for Quantum Technologies, National University of Singapore, 3 Science Drive 2, S117543, Singapore\\
}%
\affiliation{Now at Department of Electrical and Electronics Engineering, \"{O}zye\v{g}in University, 34794, Istanbul, Turkey}

\author{Christoph F. Wildfeuer}%
\affiliation{FHNW University of Applied Sciences and Arts Northwestern Switzerland, School of Engineering, Klosterzelgstrasse 2, CH-5210 Windisch, Switzerland\\
}
\author{Douglas Griffin}%
\affiliation{University of New South Wales Canberra, School of Engineering and Information Technology, Canberra, Australia\\
}
\author{Daniel K. L. Oi}%
\affiliation{SUPA Department of Physics, University of Strathclyde, John Anderson Building, 107 Rottenrow East, G4 0NG Glasgow, UK\\
}
\author{Alexander Ling}
\affiliation{%
 Centre for Quantum Technologies, National University of Singapore, 3 Science Drive 2, S117543, Singapore\\
}%
\affiliation{Physics Department, National University of Singapore, 2 Science Drive 3, S117542, Singapore}

\date{\today}

\begin{abstract}
Global quantum networks for secure communication can be realised using
large fleets of satellites distributing entangled photon-pairs between
ground-based nodes. Because the cost of a satellite depends on
its size, the smallest satellites will be most cost-effective. This
paper describes a miniaturised, polarization entangled, photon-pair
source operating on board a nano-satellite. The source violates Bell’s
inequality with a CHSH parameter of \SI[separate-uncertainty = true]{2.60(6)}{}. This source can be
combined with optical link technologies to enable future quantum
communication nano-satellite missions.   
\end{abstract}
\maketitle

\section{Introduction}

Quantum entanglement describes non-local correlation between multiple bodies such that their wavefunction is irreducible to a product of individual wavefunctions. Entanglement correlations~\cite{epr,scat,bellinequality, freedman72, aspect81} have emerged as an essential resource in quantum technologies and entanglement is used in various fields such as computation~\cite{loss1998quantum}, sensing~\cite{degen2017quantum} and communication~\cite{E91}.

Near term applications include quantum key distribution (QKD), where entanglement can be used to quantify knowledge gained by an adversary~\cite{E91} and to enable device-independent encryption~\cite{acin07}.
Beyond the immediate advantages for near-term technologies (such as QKD), efforts and resources are being directed towards the development of a Quantum Internet~\cite{kimble08,wehner2018quantum}. 
Such a network is envisioned to feature quantum nodes that are capable of producing, detecting or verifying quantum entanglement. A global quantum network can be more readily realized using space-based nodes~\cite{wehner2018quantum,vergoossen2019satellite,khatri2019spooky}.

\begin{figure}[h]
	\centering
	\includegraphics[scale=1]{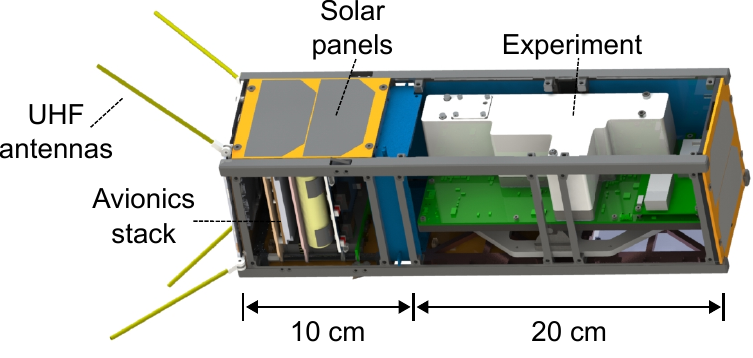}
	\caption{Systems within the SpooQy-1 satellite (some solar panels not shown for clarity). Fully assembled, the CubeSat mass is \SI{2.6}{\kilogram}. The experiment has a volume of \SI{206}{\milli\metre}x \SI{85}{\milli\metre} x \SI{49}{\milli\metre}, a mass of \SI{0.9}{\kilogram} and its peak power consumption is \SI{2.5}{\watt}.}
	\label{fig:set1}
\end{figure}

The first steps towards space-based nodes have been taken~\cite{vallone15,tang2016generation,gunthner17,yin17,liao2017satellite,takenaka17} and major milestones that demonstrate space-based quantum communication primitives have been achieved by the Micius satellite~\cite{liao2017satellite,yin17,liao18}. These pioneering space experiments used sizable satellite platforms with significant resources; the satellite mass in the space-based communication experiments have ranged between \SI{50}{\kilogram}~\cite{takenaka17} to \SI{600}{\kilogram}~\cite{liao2017satellite}. 
One opportunity for accelerating progress within the field is to utilise smaller, standardized spacecraft to enable cost-effective quantum nodes in space~\cite{morong2012quantum,oi2017cubesat}.
The de-facto standard spacecraft in the nano-satellite class is the CubeSat, where the most basic platform is a \SI{10}{\centi\metre} cube (1U) and a growing family of proportionally larger (2U, 3U, 6U, 12U) spacecraft are also defined.

Previous work reported that photon-pairs created by spontaneous parametric downconversion (SPDC) could be generated on board CubeSats~\cite{tang2016generation}. This paper reports on the essential next step: the generation and detection of polarization entangled photon-pairs on board a CubeSat in low-Earth orbit (LEO).
This demonstration marks a milestone towards realizing space-to-ground entanglement distribution from a CubeSat~\cite{qkdqubesat}.

\begin{figure}[b]
	\centering
	\includegraphics{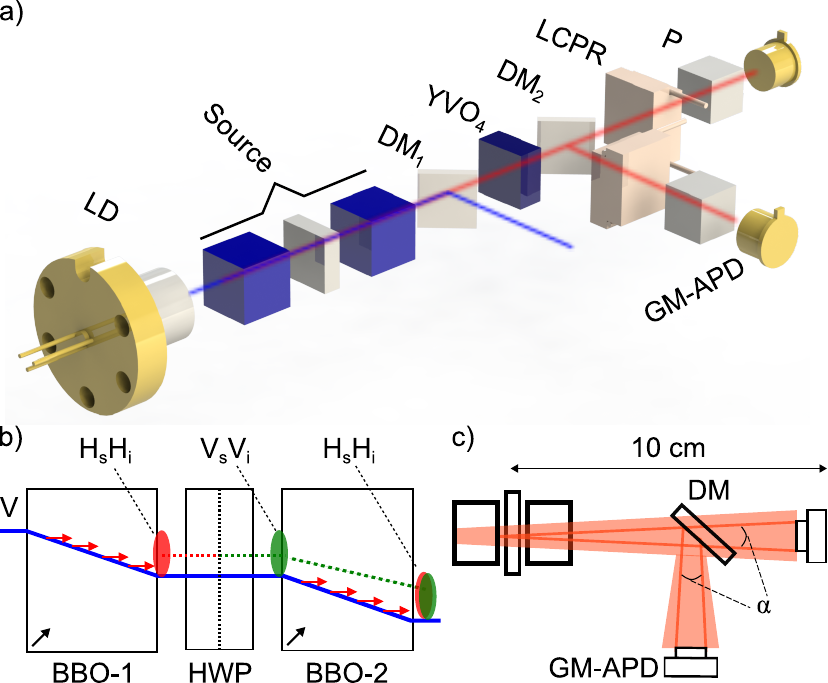}
	\caption{(a) Essential optics in the experiment. LD: laser diode; DM: dichroic mirror; YVO$_4$: yttrium orthovanadate; LCPR: liquid-crystal polarisation rotator; P: polarizer; GM-APD: Geiger-mode avalanche-photodiode. (b) Entangled photon-pair source concept. This design facilitates either of the $\ket{\Phi^{\pm}}$ Bell states. (c) Collection condition of the SPDC photons (opening angle, $\alpha$, $0.3 \SI{}{\degree}$) detected by GM-APDs.}
	\label{fig:set2}
\end{figure}

\section{Methods}
The in-orbit experiment occupies approximately 2U of volume in the 3U CubeSat, SpooQy-1 (Fig.~\ref{fig:set1}, designed and built at the Centre for Quantum Technologies, National University of Singapore). The remaining 1U houses the spacecraft avionics. The experiment is composed of a source of entangled photon-pairs coupled to a detector module (see Fig.~\ref{fig:set2}(a)) all controlled by an integrated electronics sub-system. A micro-controller on the experiment interfaces to the satellite's on-board computer to receive commands and to return science data to ground control.

The polarization entangled photon-pair source is based on collinear, non-degenerate type-I SPDC with critically phase-matched non-linear crystals. The source design (Fig.~\ref{fig:set2}(b)) uses a \textit{parallel-crystal} configuration~\cite{villar2018experimental, lohrmann2018high}. The beam overlap found in this design provides the source with better alignment stability in contrast to other two-crystal designs~\cite{trojek2008collinear}.

A collimated laser diode (central wavelength $\lambda$ = \SI{405}{\nano\metre}, spectral linewidth $\Delta\nu$=\SI{160}{\mega\hertz}) with a beam full-width half-maximum of \SI{800}{\micro\metre}$\times$\SI{400}{\micro\metre} is used as a continuous-wave pump for the SPDC process. The pump produces horizontally polarized photon-pairs in two $\beta$-Barium Borate (BBO-1 and BBO-2) crystals (cut angle: \SI{28.8}{\degree}, length: \SI{6}{\milli\metre}). Between the two BBO crystals, an achromatic half-wave plate (HWP) induces a \SI{90}{\degree} rotation in the polarization of the SPDC photons from BBO-1, while the pump polarization remains unaffected.

The photon-pair source produces the state $\ket{\phi} = \frac{1}{\sqrt{2}}\big(\ket{H_{s}H_{i}}+e^{i\Delta\varphi}\ket{V_{s}V_{i}}\big)$, where $s$ ($i$) denotes the signal (idler) photon wavelength, and $\Delta\varphi$ is the relative phase-difference between photon-pairs born in BBO-1 and BBO-2. Excess pump light is removed by a dichroic mirror to a detector that tracks power and pointing. An a-cut yttrium orthovanadate (YVO$_4$) crystal compensates for the birefringent dispersion of the SPDC photons (related to $\Delta\varphi$~\cite{villar2018experimental}). The tilt angle of BBO-1 is adjusted such that the final phase difference $\Delta\varphi$ becomes $\pi$ generating the maximally entangled Bell state $\vert \Phi^-\rangle$.

The relative angle of the pump beam and the optical axis of the BBO crystals must be kept within \SI{100}{\micro \radian} in order to control the phase of the generated photon-pairs (see Fig.~\ref{fig:SI-1} in Appendix~\ref{appendix:tolerance}). This can be achieved without active alignment using titanium flexure stages. To reduce misalignments resulting from a mismatch in the thermal expansion of different materials the rest of the optical bench is also made of titanium.

The SPDC photon-pairs are separated by a dichroic mirror, and signal and idler photons have their polarization state analysed separately. 
Each polarization analyser is composed of a liquid-crystal polarization rotator (LCPR) followed by a polarizer~\cite{lohrmann2019manipulation}.
Photon detection is performed using un-cooled, passively-quenched, Geiger-mode avalanche photodiodes (GM-APDs, with detection efficiencies of 45\% at $\SI{800}{\nano\metre}$) with active areas of $\SI{500}{\micro\metre}$ located $\SI{10}{cm}$ away from the centre of the source. 
Detection events are identified as correlations if they occur within a time window of \SI[separate-uncertainty = true]{4.84(6)}{}~ns.

To simplify the optical assembly, collection optics were not used. This collection condition, described in Fig.~\ref{fig:set2}(c), restricts the light detection to SPDC photons with an opening angle, $\alpha$, of $\SI{0.3}{\degree}$. While this affects the brightness, it does not detract from the primary objective of demonstrating in-orbit entanglement.

During the course of operation in orbit the satellite experiences varying levels of solar illumination causing the temperature of the experiment to fluctuate.
This can be mitigated by running a \SI{2.5}{\watt} heater (see Fig.~\ref{fig:SI-2} in Appendix~\ref{appendix:temps}). 
The temperature variation affects the breakdown voltage of the GM-APDs. To ensure a constant detection efficiency, the bias voltage of the detectors are optimized by a window comparator technique that tracks changes in the output pulse height of the GM-APDs~\cite{sachidananda2011bias}.

To investigate the  polarization correlation of the photon-pairs, one arm is analysed with fixed polarization (either H: horizontal, V: vertical, D: diagonal, or A: anti-diagonal) while the other arm is swept through different polarization states. In principle, the LCPR devices can achieve almost 2$\pi$ of phase shift, but towards the end of the range the performance of the devices lack precision. 
To improve performance reliability, the LCPR devices were restricted to a phase shift of approximately 150 degrees.

\begin{figure}[b]
	\centering
	\includegraphics{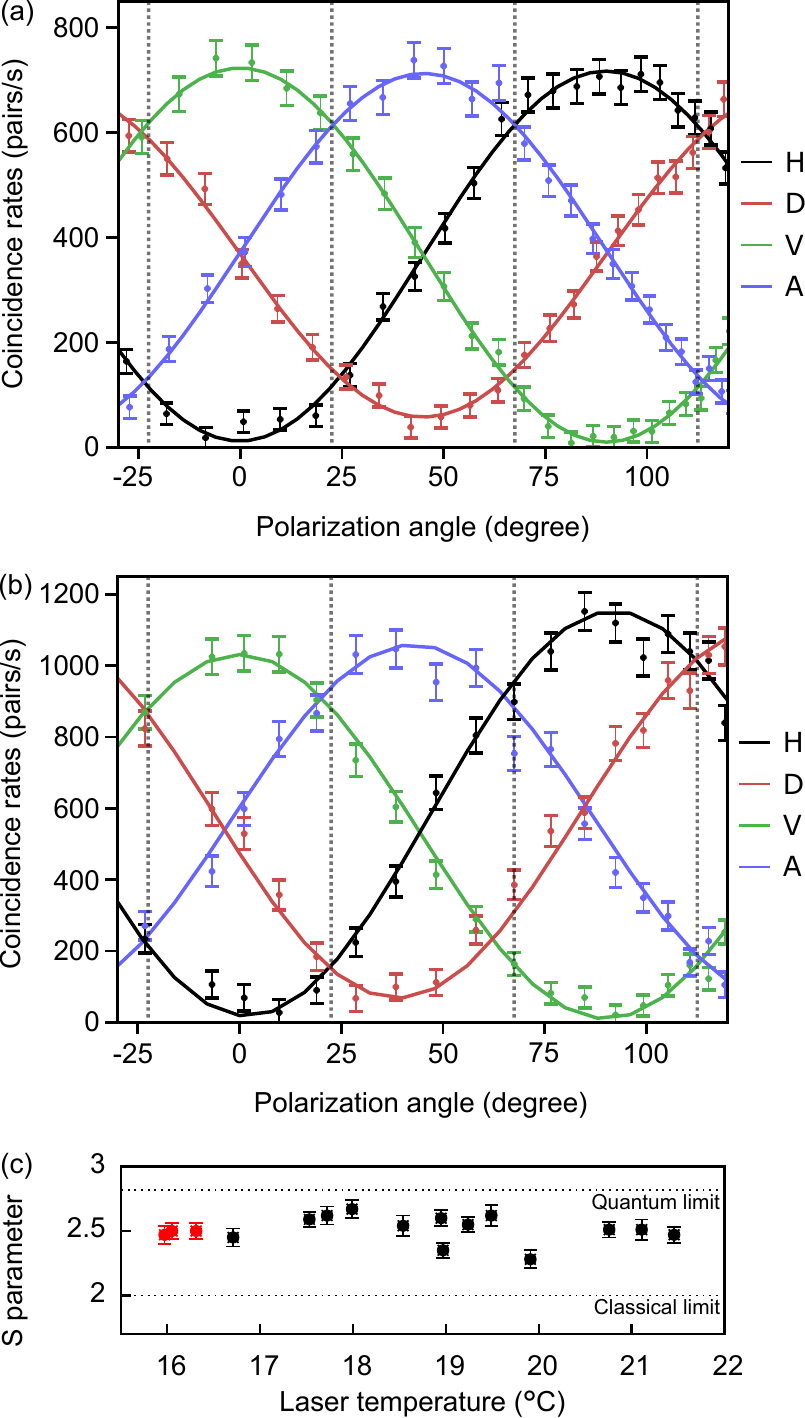}
	\caption{(a) On-ground polarization correlation curves after the experiment was integrated into the satellite. Average visibility at \SI{20}{\degree C} (corrected for accidental coincidences) recorded in the two bases ($H$/$V$ and $D$/$A$) were: $V_{H/V}$=\SI[separate-uncertainty = true]{0.97(5)}{} and $V_{D/A}$=\SI[separate-uncertainty = true]{0.87(5)}{}. The corresponding CHSH parameter was \SI[separate-uncertainty = true]{2.63(7)}{}. The dashed, vertical lines indicate the settings used to obtained the CHSH parameter. (b) Correlation curves measured in orbit on 16th July 2019. For clarity, only a subset of data points are shown. The average visibility (corrected for accidental coincidences) at \SI{17.5}{\degree C} was: $V_{H/V}$=\SI[separate-uncertainty = true]{0.98(6)}{} and $V_{D/A}$=\SI[separate-uncertainty = true]{0.88(6)}{}. The corresponding CHSH parameter of \SI[separate-uncertainty = true]{2.60(6)}{}. (c) In-orbit CHSH values at different temperatures obtained over two weeks of operation. The red-coloured data points were taken after the satellite was under direct solar illumination for 100 hours (see Fig.~\ref{fig:SI-2} in Appendix~\ref{appendix:temps}).}
	\label{fig:set3}
\end{figure}

The visibility (contrast) of the polarization correlation curves can be used to assess the quality of the entangled state. 
Additionally, it is possible to extract 16 data points from these correlation curves. Each curve can provide four data points that are separated by \SI{45}{\degree} (see Fig.~\ref{fig:set3}). These data points are used to obtain a measure of entanglement known as the Clauser-Horne-Shimony-Holt (CHSH)~\cite{clauser1969proposed} parameter, $S$.

After assembly of the satellite, the on-ground detected pair rate (combined for both polarization bases) is \SI[per-mode=symbol]{1400}{pairs/s} at approximately \SI{17}{\milli\watt} of pump power ($\approx$~\SI[per-mode=symbol]{590000}{singles/s} for signal and idler). 
The visibilities (corrected for accidentals) recorded in the two bases ($H$/$V$ and $D$/$A$) were: $V_H$=\SI[separate-uncertainty = true]{0.97(5)}{}, $V_V$=\SI[separate-uncertainty = true]{0.97(6)}{}, $V_D$=\SI[separate-uncertainty = true]{0.84(5)}{} and $V_A$=\SI[separate-uncertainty = true]{0.90(5)}{}. From these curves, a CHSH parameter of \SI[separate-uncertainty = true]{2.63(7)}{} was extracted (see Fig.~\ref{fig:set3}(a)). If used for quantum communication, this source would introduce an intrinsic quantum bit error ratio (QBER) of approximately \SI[separate-uncertainty = true]{3.9(4)}{}\%.

\section{Results}
The satellite was deployed into orbit from the International Space station on 17th June 2019 (orbit inclination: \SI{51.6}{\degree}, \SI{408}{\kilo\metre} altitude) and operations began on the same day.
The temperature of the experiment fluctuated according to the diurnal cycle of the satellite's 90 minute orbital period as expected. To bring the experiment within the range of operating temperature, the on-board heater was activated (see Fig.~\ref{fig:SI-2} in Appendix~\ref{appendix:temps}). The difference between the on-ground and in-orbit temperatures made it necessary to re-calibrate the LCPR devices and to operate the experiment in-orbit with a different pump current (see Fig.~\ref{fig:SI-3} in Appendix~\ref{appendix:heatmap}), which can yield a different laser mode leading to slightly different brightness levels.

The typical in-orbit detected pair rate (combined for both polarization bases) was \SI[per-mode=symbol]{2200}{pairs/s} ($\approx$~\SI[per-mode=symbol]{700000}{singles/s} for signal and idler). 
The highest recorded visibilities were: $V_H$=\SI[separate-uncertainty = true]{0.98(5)}{}, $V_V$=\SI[separate-uncertainty = true]{0.97(6)}{}, $V_D$=\SI[separate-uncertainty = true]{0.88(6)}{} and $V_A$=\SI[separate-uncertainty = true]{0.88(6)}{}. These visibilities yielded a CHSH parameter of 
$V_H$=\SI[separate-uncertainty = true]{2.60(6)}{} (Fig.~\ref{fig:set3}(b)).
This value is a slight underestimate of the actual CHSH parameter because the LCPR settings for the diagonal and anti-diagonal polarization states had a systematic error. 
This can be seen from Fig.~\ref{fig:set3}(b) where the extrema of the correlation curves in the diagonal/anti-diagonal settings do not occur exactly at the $D/A$ (\SI{45}{\degree}/\SI{135}{\degree}) basis setting. Nevertheless, this causes only a slight degradation in the CHSH value compared to the on-ground baseline value.

Entangled photon-pair production was observed over a temperature range from \SI{16}{\degree C} to \SI{21.5}{\degree C} (Fig.~\ref{fig:set3}(c)).
The experiment experienced relatively high temperatures when the satellite entered an orbital condition of continuous solar illumination 
(no data was collected during this period).
Data collection resumed after exiting continuous illumination and pre-illumination performance was observed (see red data points in Fig.~\ref{fig:set3}(c)).

\section{Discussion and conclusion}
The operation of a polarization entangled photon-pair source on board a CubeSat in LEO has been reported. This shows that entanglement technology can be deployed with minimal resources in novel operating environments, providing valuable `space heritage' for different components and assembling techniques.

The next generation of the experiment can achieve an improvement of two orders of magnitude in the photon-pair rate~\cite{lohrmann2018high}, and other SPDC configurations are under consideration to enable other performance improvements~\cite{lohrmanfeaturedpaper}. 
A follow-on mission is under development where the goal is to share entanglement between a nano-satellite and a ground receiver~\cite{qkdqubesat}.
To achieve this goal, it is necessary to equip a nano-satellite with an optical terminal that has a pointing capability of approximately \SI{10}{\micro rad}~\cite{oi2017cubesat}. While this additional infrastructure is demanding, solutions have been reported from the commercial sector~\cite{storm2017cubesat}.

The result from this in-orbit experiment paves the way for testing a variety of satellite-based quantum communication protocols using small standardised spacecraft such as CubeSats.
These include placement of sources of faint laser pulses in space to perform decoy-state QKD, or to install only quantum receivers on the CubeSats to enable an uplink configuration~\cite{jennewein2014nanoqey,kerstel2018nanobob,haber2018qube}.
Beyond securing keys, two-way entanglement distribution can also enable secure time-transfer~\cite{lee19} between satellites providing global navigation services.
Such a capability can be demonstrated via inter-CubeSat quantum communication~\cite{naughton2019design}.

Miniaturized sources are not restricted to nano-satellites. They can also be useful for the development of quantum communication
subsystems in larger spacecrafts. This manuscript also shows that CubeSats are well-placed to perform in-orbit subsystem and device performance
characterization to support the development of space missions with larger satellites.

As small standardized spacecraft are cost-effective, we anticipate an acceleration of space-based demonstration for quantum technologies in domains such as time-keeping and sensing.
The use of a standardized platform makes it easier to work towards a realistic miniaturization of the required technology, and to scale up the number of space-based nodes with constellations of \textit{quantum} nano-satellites to enable a global Quantum Internet.

\section{acknowledgments}
This program was supported by the National Research Foundation (Award No. NRF-CRP12-2013-02), Prime Minister’s Office of Singapore, and the Ministry of Education, Singapore. The experiment and satellite was designed and built at the Centre for Quantum Technologies, National University of Singapore.

Daniel K. L. Oi acknowledges support from the UK Space Agency (NSTP3-FT-063, NSTP-QSTP), Innovate UK (AQKD), and EPSRC (Quantum Technology Hub in Quantum Communications Partnership Resource).
\clearpage
\bibliographystyle{apsrev4-2}
\bibliography{bibliography}

\begin{thebibliography}{39}%
\makeatletter
\providecommand \@ifxundefined [1]{%
 \@ifx{#1\undefined}
}%
\providecommand \@ifnum [1]{%
 \ifnum #1\expandafter \@firstoftwo
 \else \expandafter \@secondoftwo
 \fi
}%
\providecommand \@ifx [1]{%
 \ifx #1\expandafter \@firstoftwo
 \else \expandafter \@secondoftwo
 \fi
}%
\providecommand \natexlab [1]{#1}%
\providecommand \enquote  [1]{``#1''}%
\providecommand \bibnamefont  [1]{#1}%
\providecommand \bibfnamefont [1]{#1}%
\providecommand \citenamefont [1]{#1}%
\providecommand \href@noop [0]{\@secondoftwo}%
\providecommand \href [0]{\begingroup \@sanitize@url \@href}%
\providecommand \@href[1]{\@@startlink{#1}\@@href}%
\providecommand \@@href[1]{\endgroup#1\@@endlink}%
\providecommand \@sanitize@url [0]{\catcode `\\12\catcode `\$12\catcode
  `\&12\catcode `\#12\catcode `\^12\catcode `\_12\catcode `\%12\relax}%
\providecommand \@@startlink[1]{}%
\providecommand \@@endlink[0]{}%
\providecommand \url  [0]{\begingroup\@sanitize@url \@url }%
\providecommand \@url [1]{\endgroup\@href {#1}{\urlprefix }}%
\providecommand \urlprefix  [0]{URL }%
\providecommand \Eprint [0]{\href }%
\providecommand \doibase [0]{https://doi.org/}%
\providecommand \selectlanguage [0]{\@gobble}%
\providecommand \bibinfo  [0]{\@secondoftwo}%
\providecommand \bibfield  [0]{\@secondoftwo}%
\providecommand \translation [1]{[#1]}%
\providecommand \BibitemOpen [0]{}%
\providecommand \bibitemStop [0]{}%
\providecommand \bibitemNoStop [0]{.\EOS\space}%
\providecommand \EOS [0]{\spacefactor3000\relax}%
\providecommand \BibitemShut  [1]{\csname bibitem#1\endcsname}%
\let\auto@bib@innerbib\@empty
\bibitem [{\citenamefont {Einstein}\ \emph {et~al.}(1935)\citenamefont
  {Einstein}, \citenamefont {Podolsky},\ and\ \citenamefont {Rosen}}]{epr}%
  \BibitemOpen
  \bibfield  {author} {\bibinfo {author} {\bibfnamefont {A.}~\bibnamefont
  {Einstein}}, \bibinfo {author} {\bibfnamefont {B.}~\bibnamefont {Podolsky}},\
  and\ \bibinfo {author} {\bibfnamefont {N.}~\bibnamefont {Rosen}},\ }\href
  {http://dx.doi.org/10.1103/PhysRev.47.777} {\bibfield  {journal} {\bibinfo
  {journal} {Phys. Rev.}\ }\textbf {\bibinfo {volume} {47}},\ \bibinfo {pages}
  {777–780} (\bibinfo {year} {1935})}\BibitemShut {NoStop}%
\bibitem [{\citenamefont {Schr\"{o}dinger}(1935)}]{scat}%
  \BibitemOpen
  \bibfield  {author} {\bibinfo {author} {\bibfnamefont {E.}~\bibnamefont
  {Schr\"{o}dinger}},\ }\href
  {https://link.springer.com/article/10.1007/BF01491987} {\bibfield  {journal}
  {\bibinfo  {journal} {Sci. Nat.}\ }\textbf {\bibinfo {volume} {23}},\
  \bibinfo {pages} {844–849} (\bibinfo {year} {1935})}\BibitemShut {NoStop}%
\bibitem [{\citenamefont {Bell}(1964)}]{bellinequality}%
  \BibitemOpen
  \bibfield  {author} {\bibinfo {author} {\bibfnamefont {J.~S.}\ \bibnamefont
  {Bell}},\ }\href
  {https://journals.aps.org/ppf/abstract/10.1103/PhysicsPhysiqueFizika.1.195}
  {\bibfield  {journal} {\bibinfo  {journal} {Physics}\ }\textbf {\bibinfo
  {volume} {1}},\ \bibinfo {pages} {195–200} (\bibinfo {year}
  {1964})}\BibitemShut {NoStop}%
\bibitem [{\citenamefont {Freedman}\ and\ \citenamefont
  {Clauser}(1972)}]{freedman72}%
  \BibitemOpen
  \bibfield  {author} {\bibinfo {author} {\bibfnamefont {S.~J.}\ \bibnamefont
  {Freedman}}\ and\ \bibinfo {author} {\bibfnamefont {J.~F.}\ \bibnamefont
  {Clauser}},\ }\href {https://doi.org/10.1103/PhysRevLett.28.938} {\bibfield
  {journal} {\bibinfo  {journal} {Phys. Rev. Lett.}\ }\textbf {\bibinfo
  {volume} {28}},\ \bibinfo {pages} {938–941} (\bibinfo {year}
  {1972})}\BibitemShut {NoStop}%
\bibitem [{\citenamefont {Aspect}\ \emph {et~al.}(1981)\citenamefont {Aspect},
  \citenamefont {Grangier},\ and\ \citenamefont {Roger}}]{aspect81}%
  \BibitemOpen
  \bibfield  {author} {\bibinfo {author} {\bibfnamefont {A.}~\bibnamefont
  {Aspect}}, \bibinfo {author} {\bibfnamefont {P.}~\bibnamefont {Grangier}},\
  and\ \bibinfo {author} {\bibfnamefont {G.}~\bibnamefont {Roger}},\ }\href
  {https://doi.org/10.1103/PhysRevLett.47.460} {\bibfield  {journal} {\bibinfo
  {journal} {Phys. Rev. Lett.}\ }\textbf {\bibinfo {volume} {47}},\ \bibinfo
  {pages} {460–463} (\bibinfo {year} {1981})}\BibitemShut {NoStop}%
\bibitem [{\citenamefont {Loss}\ and\ \citenamefont
  {DiVincenzo}(1998)}]{loss1998quantum}%
  \BibitemOpen
  \bibfield  {author} {\bibinfo {author} {\bibfnamefont {D.}~\bibnamefont
  {Loss}}\ and\ \bibinfo {author} {\bibfnamefont {D.~P.}\ \bibnamefont
  {DiVincenzo}},\ }\href
  {https://journals.aps.org/pra/abstract/10.1103/PhysRevA.57.120} {\bibfield
  {journal} {\bibinfo  {journal} {Phys. Rev. A}\ }\textbf {\bibinfo {volume}
  {57}},\ \bibinfo {pages} {120–126} (\bibinfo {year} {1998})}\BibitemShut
  {NoStop}%
\bibitem [{\citenamefont {Degen}\ \emph {et~al.}(2017)\citenamefont {Degen},
  \citenamefont {Reinhard},\ and\ \citenamefont
  {Cappellaro}}]{degen2017quantum}%
  \BibitemOpen
  \bibfield  {author} {\bibinfo {author} {\bibfnamefont {C.~L.}\ \bibnamefont
  {Degen}}, \bibinfo {author} {\bibfnamefont {F.}~\bibnamefont {Reinhard}},\
  and\ \bibinfo {author} {\bibfnamefont {P.}~\bibnamefont {Cappellaro}},\
  }\href {https://journals.aps.org/rmp/abstract/10.1103/RevModPhys.89.035002}
  {\bibfield  {journal} {\bibinfo  {journal} {Rev. Mod. Phys.}\ }\textbf
  {\bibinfo {volume} {89}},\ \bibinfo {pages} {035002} (\bibinfo {year}
  {2017})}\BibitemShut {NoStop}%
\bibitem [{\citenamefont {Ekert}(1991)}]{E91}%
  \BibitemOpen
  \bibfield  {author} {\bibinfo {author} {\bibfnamefont {A.~K.}\ \bibnamefont
  {Ekert}},\ }\href {https://doi.org/10.1103/PhysRevLett.67.661} {\bibfield
  {journal} {\bibinfo  {journal} {Phys. Rev. Lett.}\ }\textbf {\bibinfo
  {volume} {67}},\ \bibinfo {pages} {661–663} (\bibinfo {year}
  {1991})}\BibitemShut {NoStop}%
\bibitem [{\citenamefont {Acin}\ \emph {et~al.}(2007)\citenamefont {Acin},
  \citenamefont {Brunner}, \citenamefont {Gisin}, \citenamefont {Massar},
  \citenamefont {Pironio},\ and\ \citenamefont {Scarani}}]{acin07}%
  \BibitemOpen
  \bibfield  {author} {\bibinfo {author} {\bibfnamefont {A.}~\bibnamefont
  {Acin}}, \bibinfo {author} {\bibfnamefont {N.}~\bibnamefont {Brunner}},
  \bibinfo {author} {\bibfnamefont {N.}~\bibnamefont {Gisin}}, \bibinfo
  {author} {\bibfnamefont {S.}~\bibnamefont {Massar}}, \bibinfo {author}
  {\bibfnamefont {S.}~\bibnamefont {Pironio}},\ and\ \bibinfo {author}
  {\bibfnamefont {V.}~\bibnamefont {Scarani}},\ }\href
  {https://doi.org/10.1103/PhysRevLett.98.230501} {\bibfield  {journal}
  {\bibinfo  {journal} {Phys.Rev. Lett.}\ }\textbf {\bibinfo {volume} {98}},\
  \bibinfo {eid} {230501} (\bibinfo {year} {2007})}\BibitemShut {NoStop}%
\bibitem [{\citenamefont {Kimble}(2008)}]{kimble08}%
  \BibitemOpen
  \bibfield  {author} {\bibinfo {author} {\bibfnamefont {H.~J.}\ \bibnamefont
  {Kimble}},\ }\href {https://www.nature.com/articles/nature07127} {\bibfield
  {journal} {\bibinfo  {journal} {Nature}\ }\textbf {\bibinfo {volume} {453}},\
  \bibinfo {pages} {1023–1030} (\bibinfo {year} {2008})}\BibitemShut
  {NoStop}%
\bibitem [{\citenamefont {Wehner}\ \emph {et~al.}(2018)\citenamefont {Wehner},
  \citenamefont {Elkouss},\ and\ \citenamefont {Hanson}}]{wehner2018quantum}%
  \BibitemOpen
  \bibfield  {author} {\bibinfo {author} {\bibfnamefont {S.}~\bibnamefont
  {Wehner}}, \bibinfo {author} {\bibfnamefont {D.}~\bibnamefont {Elkouss}},\
  and\ \bibinfo {author} {\bibfnamefont {R.}~\bibnamefont {Hanson}},\ }\href
  {https://science.sciencemag.org/content/362/6412/eaam9288} {\bibfield
  {journal} {\bibinfo  {journal} {Science}\ }\textbf {\bibinfo {volume} {362}}
  (\bibinfo {year} {2018})}\BibitemShut {NoStop}%
\bibitem [{\citenamefont {Vergoossen}\ \emph {et~al.}(2020)\citenamefont
  {Vergoossen}, \citenamefont {Loarte}, \citenamefont {Bedington},
  \citenamefont {Kuiper},\ and\ \citenamefont
  {Ling}}]{vergoossen2019satellite}%
  \BibitemOpen
  \bibfield  {author} {\bibinfo {author} {\bibfnamefont {T.}~\bibnamefont
  {Vergoossen}}, \bibinfo {author} {\bibfnamefont {S.}~\bibnamefont {Loarte}},
  \bibinfo {author} {\bibfnamefont {R.}~\bibnamefont {Bedington}}, \bibinfo
  {author} {\bibfnamefont {H.}~\bibnamefont {Kuiper}},\ and\ \bibinfo {author}
  {\bibfnamefont {A.}~\bibnamefont {Ling}},\ }\href
  {https://doi.org/https://doi.org/10.1016/j.actaastro.2020.02.010} {\bibfield
  {journal} {\bibinfo  {journal} {Acta Astronautica}\ }\textbf {\bibinfo
  {volume} {0094-5765}},\ \bibinfo {pages} {1–12} (\bibinfo {year}
  {2020})}\BibitemShut {NoStop}%
\bibitem [{\citenamefont {Khatri}\ \emph {et~al.}(2019)\citenamefont {Khatri},
  \citenamefont {Brady}, \citenamefont {Desporte}, \citenamefont {Bart},\ and\
  \citenamefont {Dowling}}]{khatri2019spooky}%
  \BibitemOpen
  \bibfield  {author} {\bibinfo {author} {\bibfnamefont {S.}~\bibnamefont
  {Khatri}}, \bibinfo {author} {\bibfnamefont {A.~J.}\ \bibnamefont {Brady}},
  \bibinfo {author} {\bibfnamefont {R.~A.}\ \bibnamefont {Desporte}}, \bibinfo
  {author} {\bibfnamefont {M.~P.}\ \bibnamefont {Bart}},\ and\ \bibinfo
  {author} {\bibfnamefont {J.~P.}\ \bibnamefont {Dowling}},\ }\href
  {https://arxiv.org/pdf/1912.06678.pdf} {\bibfield  {journal} {\bibinfo
  {journal} {arXiv preprint arXiv:1912.06678}\ } (\bibinfo {year}
  {2019})}\BibitemShut {NoStop}%
\bibitem [{\citenamefont {Vallone}\ \emph {et~al.}(2015)\citenamefont
  {Vallone}, \citenamefont {Bacco}, \citenamefont {Dequal}, \citenamefont
  {Gaiarin}, \citenamefont {Luceri}, \citenamefont {Bianco},\ and\
  \citenamefont {Villoresi}}]{vallone15}%
  \BibitemOpen
  \bibfield  {author} {\bibinfo {author} {\bibfnamefont {G.}~\bibnamefont
  {Vallone}}, \bibinfo {author} {\bibfnamefont {D.}~\bibnamefont {Bacco}},
  \bibinfo {author} {\bibfnamefont {D.}~\bibnamefont {Dequal}}, \bibinfo
  {author} {\bibfnamefont {S.}~\bibnamefont {Gaiarin}}, \bibinfo {author}
  {\bibfnamefont {V.}~\bibnamefont {Luceri}}, \bibinfo {author} {\bibfnamefont
  {G.}~\bibnamefont {Bianco}},\ and\ \bibinfo {author} {\bibfnamefont
  {P.}~\bibnamefont {Villoresi}},\ }\href
  {https://doi.org/10.1103/PhysRevLett.115.040502} {\bibfield  {journal}
  {\bibinfo  {journal} {Phys. Rev. Lett.}\ }\textbf {\bibinfo {volume} {115}},\
  \bibinfo {pages} {040502} (\bibinfo {year} {2015})}\BibitemShut {NoStop}%
\bibitem [{\citenamefont {Tang}\ \emph {et~al.}(2016)\citenamefont {Tang},
  \citenamefont {Chandrasekara}, \citenamefont {Tan}, \citenamefont {Cheng},
  \citenamefont {Sha}, \citenamefont {Hiang}, \citenamefont {Oi},\ and\
  \citenamefont {Ling}}]{tang2016generation}%
  \BibitemOpen
  \bibfield  {author} {\bibinfo {author} {\bibfnamefont {Z.}~\bibnamefont
  {Tang}}, \bibinfo {author} {\bibfnamefont {R.}~\bibnamefont {Chandrasekara}},
  \bibinfo {author} {\bibfnamefont {Y.~C.}\ \bibnamefont {Tan}}, \bibinfo
  {author} {\bibfnamefont {C.}~\bibnamefont {Cheng}}, \bibinfo {author}
  {\bibfnamefont {L.}~\bibnamefont {Sha}}, \bibinfo {author} {\bibfnamefont
  {G.~C.}\ \bibnamefont {Hiang}}, \bibinfo {author} {\bibfnamefont {D.~K.}\
  \bibnamefont {Oi}},\ and\ \bibinfo {author} {\bibfnamefont {A.}~\bibnamefont
  {Ling}},\ }\href
  {https://journals.aps.org/prapplied/abstract/10.1103/PhysRevApplied.5.054022}
  {\bibfield  {journal} {\bibinfo  {journal} {Phys. Rev. Appl.}\ }\textbf
  {\bibinfo {volume} {5}},\ \bibinfo {pages} {054022} (\bibinfo {year}
  {2016})}\BibitemShut {NoStop}%
\bibitem [{\citenamefont {G\"{u}nthner}\ \emph {et~al.}(2017)\citenamefont
  {G\"{u}nthner}, \citenamefont {Khan}, \citenamefont {Elser}, \citenamefont
  {Stiller}, \citenamefont {\"{O}mer Bayraktar}, \citenamefont {M\"{u}ller},
  \citenamefont {Saucke}, \citenamefont {Tr\"{o}ndle}, \citenamefont {Heine},
  \citenamefont {Seel}, \citenamefont {Greulich}, \citenamefont {Zech},
  \citenamefont {G\"{u}tlich}, \citenamefont {Philipp-May}, \citenamefont
  {Marquardt},\ and\ \citenamefont {Leuchs}}]{gunthner17}%
  \BibitemOpen
  \bibfield  {author} {\bibinfo {author} {\bibfnamefont {K.}~\bibnamefont
  {G\"{u}nthner}}, \bibinfo {author} {\bibfnamefont {I.}~\bibnamefont {Khan}},
  \bibinfo {author} {\bibfnamefont {D.}~\bibnamefont {Elser}}, \bibinfo
  {author} {\bibfnamefont {B.}~\bibnamefont {Stiller}}, \bibinfo {author}
  {\bibnamefont {\"{O}mer Bayraktar}}, \bibinfo {author} {\bibfnamefont
  {C.~R.}\ \bibnamefont {M\"{u}ller}}, \bibinfo {author} {\bibfnamefont
  {K.}~\bibnamefont {Saucke}}, \bibinfo {author} {\bibfnamefont
  {D.}~\bibnamefont {Tr\"{o}ndle}}, \bibinfo {author} {\bibfnamefont
  {F.}~\bibnamefont {Heine}}, \bibinfo {author} {\bibfnamefont
  {S.}~\bibnamefont {Seel}}, \bibinfo {author} {\bibfnamefont {P.}~\bibnamefont
  {Greulich}}, \bibinfo {author} {\bibfnamefont {H.}~\bibnamefont {Zech}},
  \bibinfo {author} {\bibfnamefont {B.}~\bibnamefont {G\"{u}tlich}}, \bibinfo
  {author} {\bibfnamefont {S.}~\bibnamefont {Philipp-May}}, \bibinfo {author}
  {\bibfnamefont {C.}~\bibnamefont {Marquardt}},\ and\ \bibinfo {author}
  {\bibfnamefont {G.}~\bibnamefont {Leuchs}},\ }\href
  {https://doi.org/10.1364/OPTICA.4.000611} {\bibfield  {journal} {\bibinfo
  {journal} {Optica}\ }\textbf {\bibinfo {volume} {4}},\ \bibinfo {pages}
  {611–616} (\bibinfo {year} {2017})}\BibitemShut {NoStop}%
\bibitem [{\citenamefont {Yin}\ \emph {et~al.}(2017)\citenamefont {Yin},
  \citenamefont {Cao}, \citenamefont {Li}, \citenamefont {Liao}, \citenamefont
  {Zhang}, \citenamefont {Ren}, \citenamefont {Cai}, \citenamefont {Liu},
  \citenamefont {Li}, \citenamefont {Dai}, \citenamefont {Li}, \citenamefont
  {Lu}, \citenamefont {Gong}, \citenamefont {Xu}, \citenamefont {Li},
  \citenamefont {Li}, \citenamefont {Yin}, \citenamefont {Jiang}, \citenamefont
  {Li}, \citenamefont {Jia}, \citenamefont {Ren}, \citenamefont {He},
  \citenamefont {Zhou}, \citenamefont {Zhang}, \citenamefont {Wang},
  \citenamefont {Chang}, \citenamefont {Zhu}, \citenamefont {Liu},
  \citenamefont {Chen}, \citenamefont {Lu}, \citenamefont {Shu}, \citenamefont
  {Peng}, \citenamefont {Wang},\ and\ \citenamefont {Pan}}]{yin17}%
  \BibitemOpen
  \bibfield  {author} {\bibinfo {author} {\bibfnamefont {J.}~\bibnamefont
  {Yin}}, \bibinfo {author} {\bibfnamefont {Y.}~\bibnamefont {Cao}}, \bibinfo
  {author} {\bibfnamefont {Y.-H.}\ \bibnamefont {Li}}, \bibinfo {author}
  {\bibfnamefont {S.-K.}\ \bibnamefont {Liao}}, \bibinfo {author}
  {\bibfnamefont {L.}~\bibnamefont {Zhang}}, \bibinfo {author} {\bibfnamefont
  {J.-G.}\ \bibnamefont {Ren}}, \bibinfo {author} {\bibfnamefont {W.-Q.}\
  \bibnamefont {Cai}}, \bibinfo {author} {\bibfnamefont {W.-Y.}\ \bibnamefont
  {Liu}}, \bibinfo {author} {\bibfnamefont {B.}~\bibnamefont {Li}}, \bibinfo
  {author} {\bibfnamefont {H.}~\bibnamefont {Dai}}, \bibinfo {author}
  {\bibfnamefont {G.-B.}\ \bibnamefont {Li}}, \bibinfo {author} {\bibfnamefont
  {Q.-M.}\ \bibnamefont {Lu}}, \bibinfo {author} {\bibfnamefont {Y.-H.}\
  \bibnamefont {Gong}}, \bibinfo {author} {\bibfnamefont {Y.}~\bibnamefont
  {Xu}}, \bibinfo {author} {\bibfnamefont {S.-L.}\ \bibnamefont {Li}}, \bibinfo
  {author} {\bibfnamefont {F.-Z.}\ \bibnamefont {Li}}, \bibinfo {author}
  {\bibfnamefont {Y.-Y.}\ \bibnamefont {Yin}}, \bibinfo {author} {\bibfnamefont
  {Z.-Q.}\ \bibnamefont {Jiang}}, \bibinfo {author} {\bibfnamefont
  {M.}~\bibnamefont {Li}}, \bibinfo {author} {\bibfnamefont {J.-J.}\
  \bibnamefont {Jia}}, \bibinfo {author} {\bibfnamefont {G.}~\bibnamefont
  {Ren}}, \bibinfo {author} {\bibfnamefont {D.}~\bibnamefont {He}}, \bibinfo
  {author} {\bibfnamefont {Y.-L.}\ \bibnamefont {Zhou}}, \bibinfo {author}
  {\bibfnamefont {X.-X.}\ \bibnamefont {Zhang}}, \bibinfo {author}
  {\bibfnamefont {N.}~\bibnamefont {Wang}}, \bibinfo {author} {\bibfnamefont
  {X.}~\bibnamefont {Chang}}, \bibinfo {author} {\bibfnamefont {Z.-C.}\
  \bibnamefont {Zhu}}, \bibinfo {author} {\bibfnamefont {N.-L.}\ \bibnamefont
  {Liu}}, \bibinfo {author} {\bibfnamefont {Y.-A.}\ \bibnamefont {Chen}},
  \bibinfo {author} {\bibfnamefont {C.-Y.}\ \bibnamefont {Lu}}, \bibinfo
  {author} {\bibfnamefont {R.}~\bibnamefont {Shu}}, \bibinfo {author}
  {\bibfnamefont {C.-Z.}\ \bibnamefont {Peng}}, \bibinfo {author}
  {\bibfnamefont {J.-Y.}\ \bibnamefont {Wang}},\ and\ \bibinfo {author}
  {\bibfnamefont {J.-W.}\ \bibnamefont {Pan}},\ }\href
  {https://doi.org/10.1126/science.aan3211} {\bibfield  {journal} {\bibinfo
  {journal} {Science}\ }\textbf {\bibinfo {volume} {356}},\ \bibinfo {pages}
  {1140} (\bibinfo {year} {2017})}\BibitemShut {NoStop}%
\bibitem [{\citenamefont {Liao}\ \emph {et~al.}(2017)\citenamefont {Liao},
  \citenamefont {Cai}, \citenamefont {Liu}, \citenamefont {Zhang},
  \citenamefont {Li}, \citenamefont {Ren}, \citenamefont {Yin}, \citenamefont
  {Shen}, \citenamefont {Cao}, \citenamefont {Li}, \citenamefont {Li},
  \citenamefont {Chen}, \citenamefont {Sun}, \citenamefont {Jia}, \citenamefont
  {Wu}, \citenamefont {Jiang}, \citenamefont {Wang}, \citenamefont {Huang},
  \citenamefont {Wang}, \citenamefont {Zhou}, \citenamefont {Deng},
  \citenamefont {Xi}, \citenamefont {Ma}, \citenamefont {Hu}, \citenamefont
  {Zhang}, \citenamefont {Chen}, \citenamefont {Liu}, \citenamefont {Wang},
  \citenamefont {Zhu}, \citenamefont {Lu}, \citenamefont {Shu}, \citenamefont
  {Peng}, \citenamefont {Wang},\ and\ \citenamefont {Pan}}]{liao2017satellite}%
  \BibitemOpen
  \bibfield  {author} {\bibinfo {author} {\bibfnamefont {S.-K.}\ \bibnamefont
  {Liao}}, \bibinfo {author} {\bibfnamefont {W.-Q.}\ \bibnamefont {Cai}},
  \bibinfo {author} {\bibfnamefont {W.-Y.}\ \bibnamefont {Liu}}, \bibinfo
  {author} {\bibfnamefont {L.}~\bibnamefont {Zhang}}, \bibinfo {author}
  {\bibfnamefont {Y.}~\bibnamefont {Li}}, \bibinfo {author} {\bibfnamefont
  {J.-G.}\ \bibnamefont {Ren}}, \bibinfo {author} {\bibfnamefont
  {J.}~\bibnamefont {Yin}}, \bibinfo {author} {\bibfnamefont {Q.}~\bibnamefont
  {Shen}}, \bibinfo {author} {\bibfnamefont {Y.}~\bibnamefont {Cao}}, \bibinfo
  {author} {\bibfnamefont {Z.-P.}\ \bibnamefont {Li}}, \bibinfo {author}
  {\bibfnamefont {F.-Z.}\ \bibnamefont {Li}}, \bibinfo {author} {\bibfnamefont
  {X.-W.}\ \bibnamefont {Chen}}, \bibinfo {author} {\bibfnamefont {L.-H.}\
  \bibnamefont {Sun}}, \bibinfo {author} {\bibfnamefont {J.-J.}\ \bibnamefont
  {Jia}}, \bibinfo {author} {\bibfnamefont {J.-C.}\ \bibnamefont {Wu}},
  \bibinfo {author} {\bibfnamefont {X.-J.}\ \bibnamefont {Jiang}}, \bibinfo
  {author} {\bibfnamefont {J.-F.}\ \bibnamefont {Wang}}, \bibinfo {author}
  {\bibfnamefont {Y.-M.}\ \bibnamefont {Huang}}, \bibinfo {author}
  {\bibfnamefont {Q.}~\bibnamefont {Wang}}, \bibinfo {author} {\bibfnamefont
  {Y.-L.}\ \bibnamefont {Zhou}}, \bibinfo {author} {\bibfnamefont
  {L.}~\bibnamefont {Deng}}, \bibinfo {author} {\bibfnamefont {T.}~\bibnamefont
  {Xi}}, \bibinfo {author} {\bibfnamefont {L.}~\bibnamefont {Ma}}, \bibinfo
  {author} {\bibfnamefont {T.}~\bibnamefont {Hu}}, \bibinfo {author}
  {\bibfnamefont {Q.}~\bibnamefont {Zhang}}, \bibinfo {author} {\bibfnamefont
  {Y.-A.}\ \bibnamefont {Chen}}, \bibinfo {author} {\bibfnamefont {N.-L.}\
  \bibnamefont {Liu}}, \bibinfo {author} {\bibfnamefont {X.-B.}\ \bibnamefont
  {Wang}}, \bibinfo {author} {\bibfnamefont {Z.-C.}\ \bibnamefont {Zhu}},
  \bibinfo {author} {\bibfnamefont {C.-Y.}\ \bibnamefont {Lu}}, \bibinfo
  {author} {\bibfnamefont {R.}~\bibnamefont {Shu}}, \bibinfo {author}
  {\bibfnamefont {C.-Z.}\ \bibnamefont {Peng}}, \bibinfo {author}
  {\bibfnamefont {J.-Y.}\ \bibnamefont {Wang}},\ and\ \bibinfo {author}
  {\bibfnamefont {J.-W.}\ \bibnamefont {Pan}},\ }\href
  {https://www.nature.com/articles/nature23655} {\bibfield  {journal} {\bibinfo
   {journal} {Nature}\ }\textbf {\bibinfo {volume} {549}},\ \bibinfo {pages}
  {43–47} (\bibinfo {year} {2017})}\BibitemShut {NoStop}%
\bibitem [{\citenamefont {Takenaka}\ \emph {et~al.}(2017)\citenamefont
  {Takenaka}, \citenamefont {Carrasco-Casado}, \citenamefont {Fujiwara},
  \citenamefont {Kitamura}, \citenamefont {Sasaki},\ and\ \citenamefont
  {Toyoshima}}]{takenaka17}%
  \BibitemOpen
  \bibfield  {author} {\bibinfo {author} {\bibfnamefont {H.}~\bibnamefont
  {Takenaka}}, \bibinfo {author} {\bibfnamefont {A.}~\bibnamefont
  {Carrasco-Casado}}, \bibinfo {author} {\bibfnamefont {M.}~\bibnamefont
  {Fujiwara}}, \bibinfo {author} {\bibfnamefont {M.}~\bibnamefont {Kitamura}},
  \bibinfo {author} {\bibfnamefont {M.}~\bibnamefont {Sasaki}},\ and\ \bibinfo
  {author} {\bibfnamefont {M.}~\bibnamefont {Toyoshima}},\ }\href
  {https://doi.org/10.1038/nphoton.2017.107} {\bibfield  {journal} {\bibinfo
  {journal} {Nature Photonics}\ }\textbf {\bibinfo {volume} {11}} (\bibinfo
  {year} {2017})}\BibitemShut {NoStop}%
\bibitem [{\citenamefont {Liao}\ \emph {et~al.}(2018)\citenamefont {Liao},
  \citenamefont {Cai}, \citenamefont {Handsteiner}, \citenamefont {Liu},
  \citenamefont {Yin}, \citenamefont {Zhang}, \citenamefont {Rauch},
  \citenamefont {Fink}, \citenamefont {Ren}, \citenamefont {Liu}, \citenamefont
  {Li}, \citenamefont {Shen}, \citenamefont {Cao}, \citenamefont {Li},
  \citenamefont {Wang}, \citenamefont {Huang}, \citenamefont {Deng},
  \citenamefont {Xi}, \citenamefont {Ma}, \citenamefont {Hu}, \citenamefont
  {Li}, \citenamefont {Liu}, \citenamefont {Koidl}, \citenamefont {Wang},
  \citenamefont {Chen}, \citenamefont {Wang}, \citenamefont {Steindorfer},
  \citenamefont {Kirchner}, \citenamefont {Lu}, \citenamefont {Shu},
  \citenamefont {Ursin}, \citenamefont {Scheidl}, \citenamefont {Peng},
  \citenamefont {Wang}, \citenamefont {Zeilinger},\ and\ \citenamefont
  {Pan}}]{liao18}%
  \BibitemOpen
  \bibfield  {author} {\bibinfo {author} {\bibfnamefont {S.-K.}\ \bibnamefont
  {Liao}}, \bibinfo {author} {\bibfnamefont {W.-Q.}\ \bibnamefont {Cai}},
  \bibinfo {author} {\bibfnamefont {J.}~\bibnamefont {Handsteiner}}, \bibinfo
  {author} {\bibfnamefont {B.}~\bibnamefont {Liu}}, \bibinfo {author}
  {\bibfnamefont {J.}~\bibnamefont {Yin}}, \bibinfo {author} {\bibfnamefont
  {L.}~\bibnamefont {Zhang}}, \bibinfo {author} {\bibfnamefont
  {D.}~\bibnamefont {Rauch}}, \bibinfo {author} {\bibfnamefont
  {M.}~\bibnamefont {Fink}}, \bibinfo {author} {\bibfnamefont {J.-G.}\
  \bibnamefont {Ren}}, \bibinfo {author} {\bibfnamefont {W.-Y.}\ \bibnamefont
  {Liu}}, \bibinfo {author} {\bibfnamefont {Y.}~\bibnamefont {Li}}, \bibinfo
  {author} {\bibfnamefont {Q.}~\bibnamefont {Shen}}, \bibinfo {author}
  {\bibfnamefont {Y.}~\bibnamefont {Cao}}, \bibinfo {author} {\bibfnamefont
  {F.-Z.}\ \bibnamefont {Li}}, \bibinfo {author} {\bibfnamefont {J.-F.}\
  \bibnamefont {Wang}}, \bibinfo {author} {\bibfnamefont {Y.-M.}\ \bibnamefont
  {Huang}}, \bibinfo {author} {\bibfnamefont {L.}~\bibnamefont {Deng}},
  \bibinfo {author} {\bibfnamefont {T.}~\bibnamefont {Xi}}, \bibinfo {author}
  {\bibfnamefont {L.}~\bibnamefont {Ma}}, \bibinfo {author} {\bibfnamefont
  {T.}~\bibnamefont {Hu}}, \bibinfo {author} {\bibfnamefont {L.}~\bibnamefont
  {Li}}, \bibinfo {author} {\bibfnamefont {N.-L.}\ \bibnamefont {Liu}},
  \bibinfo {author} {\bibfnamefont {F.}~\bibnamefont {Koidl}}, \bibinfo
  {author} {\bibfnamefont {P.}~\bibnamefont {Wang}}, \bibinfo {author}
  {\bibfnamefont {Y.-A.}\ \bibnamefont {Chen}}, \bibinfo {author}
  {\bibfnamefont {X.-B.}\ \bibnamefont {Wang}}, \bibinfo {author}
  {\bibfnamefont {M.}~\bibnamefont {Steindorfer}}, \bibinfo {author}
  {\bibfnamefont {G.}~\bibnamefont {Kirchner}}, \bibinfo {author}
  {\bibfnamefont {C.-Y.}\ \bibnamefont {Lu}}, \bibinfo {author} {\bibfnamefont
  {R.}~\bibnamefont {Shu}}, \bibinfo {author} {\bibfnamefont {R.}~\bibnamefont
  {Ursin}}, \bibinfo {author} {\bibfnamefont {T.}~\bibnamefont {Scheidl}},
  \bibinfo {author} {\bibfnamefont {C.-Z.}\ \bibnamefont {Peng}}, \bibinfo
  {author} {\bibfnamefont {J.-Y.}\ \bibnamefont {Wang}}, \bibinfo {author}
  {\bibfnamefont {A.}~\bibnamefont {Zeilinger}},\ and\ \bibinfo {author}
  {\bibfnamefont {J.-W.}\ \bibnamefont {Pan}},\ }\href
  {https://doi.org/10.1103/PhysRevLett.120.030501} {\bibfield  {journal}
  {\bibinfo  {journal} {Phys. Rev. Lett.}\ }\textbf {\bibinfo {volume} {120}},\
  \bibinfo {pages} {030501} (\bibinfo {year} {2018})}\BibitemShut {NoStop}%
\bibitem [{\citenamefont {Morong}\ \emph {et~al.}(2012)\citenamefont {Morong},
  \citenamefont {Ling},\ and\ \citenamefont {Oi}}]{morong2012quantum}%
  \BibitemOpen
  \bibfield  {author} {\bibinfo {author} {\bibfnamefont {W.}~\bibnamefont
  {Morong}}, \bibinfo {author} {\bibfnamefont {A.}~\bibnamefont {Ling}},\ and\
  \bibinfo {author} {\bibfnamefont {D.}~\bibnamefont {Oi}},\ }\href
  {https://www.osapublishing.org/opn/abstract.cfm?uri=opn-23-10-42} {\bibfield
  {journal} {\bibinfo  {journal} {Opt. and Photon. News}\ }\textbf {\bibinfo
  {volume} {23}},\ \bibinfo {pages} {42–49} (\bibinfo {year}
  {2012})}\BibitemShut {NoStop}%
\bibitem [{\citenamefont {Oi}\ \emph {et~al.}(2017)\citenamefont {Oi},
  \citenamefont {Ling}, \citenamefont {Vallone}, \citenamefont {Villoresi},
  \citenamefont {Greenland}, \citenamefont {Kerr}, \citenamefont {Macdonald},
  \citenamefont {Weinfurter}, \citenamefont {Kuiper}, \citenamefont {Charbon},\
  and\ \citenamefont {Ursin}}]{oi2017cubesat}%
  \BibitemOpen
  \bibfield  {author} {\bibinfo {author} {\bibfnamefont {D.~K.}\ \bibnamefont
  {Oi}}, \bibinfo {author} {\bibfnamefont {A.}~\bibnamefont {Ling}}, \bibinfo
  {author} {\bibfnamefont {G.}~\bibnamefont {Vallone}}, \bibinfo {author}
  {\bibfnamefont {P.}~\bibnamefont {Villoresi}}, \bibinfo {author}
  {\bibfnamefont {S.}~\bibnamefont {Greenland}}, \bibinfo {author}
  {\bibfnamefont {E.}~\bibnamefont {Kerr}}, \bibinfo {author} {\bibfnamefont
  {M.}~\bibnamefont {Macdonald}}, \bibinfo {author} {\bibfnamefont
  {H.}~\bibnamefont {Weinfurter}}, \bibinfo {author} {\bibfnamefont
  {H.}~\bibnamefont {Kuiper}}, \bibinfo {author} {\bibfnamefont
  {E.}~\bibnamefont {Charbon}},\ and\ \bibinfo {author} {\bibfnamefont
  {R.}~\bibnamefont {Ursin}},\ }\href
  {https://epjquantumtechnology.springeropen.com/articles/10.1140/epjqt/s40507-017-0060-1}
  {\bibfield  {journal} {\bibinfo  {journal} {EPJ Quantum Technol.}\ }\textbf
  {\bibinfo {volume} {4}},\ \bibinfo {pages} {6} (\bibinfo {year}
  {2017})}\BibitemShut {NoStop}%
\bibitem [{\citenamefont {{Centre for Quantum
  Technologies}}(2018)}]{qkdqubesat}%
  \BibitemOpen
  \bibfield  {author} {\bibinfo {author} {\bibnamefont {{Centre for Quantum
  Technologies}}},\ }\href
  {https://www.quantumlah.org/about/highlight.php?id=313} {\bibinfo {title}
  {Singapore and {U.K.} collaborate on {S}\$18m project to develop
  quantum-secured communications networks}} (\bibinfo {year}
  {2018})\BibitemShut {NoStop}%
\bibitem [{\citenamefont {Villar}\ \emph {et~al.}(2018)\citenamefont {Villar},
  \citenamefont {Lohrmann},\ and\ \citenamefont
  {Ling}}]{villar2018experimental}%
  \BibitemOpen
  \bibfield  {author} {\bibinfo {author} {\bibfnamefont {A.}~\bibnamefont
  {Villar}}, \bibinfo {author} {\bibfnamefont {A.}~\bibnamefont {Lohrmann}},\
  and\ \bibinfo {author} {\bibfnamefont {A.}~\bibnamefont {Ling}},\ }\href
  {https://www.osapublishing.org/oe/abstract.cfm?uri=oe-26-10-12396} {\bibfield
   {journal} {\bibinfo  {journal} {Opt. Express}\ }\textbf {\bibinfo {volume}
  {26}},\ \bibinfo {pages} {12396–12402} (\bibinfo {year}
  {2018})}\BibitemShut {NoStop}%
\bibitem [{\citenamefont {Lohrmann}\ \emph {et~al.}(2018)\citenamefont
  {Lohrmann}, \citenamefont {Villar}, \citenamefont {Stolk},\ and\
  \citenamefont {Ling}}]{lohrmann2018high}%
  \BibitemOpen
  \bibfield  {author} {\bibinfo {author} {\bibfnamefont {A.}~\bibnamefont
  {Lohrmann}}, \bibinfo {author} {\bibfnamefont {A.}~\bibnamefont {Villar}},
  \bibinfo {author} {\bibfnamefont {A.}~\bibnamefont {Stolk}},\ and\ \bibinfo
  {author} {\bibfnamefont {A.}~\bibnamefont {Ling}},\ }\href
  {https://aip.scitation.org/doi/abs/10.1063/1.5046962} {\bibfield  {journal}
  {\bibinfo  {journal} {Appl. Phys. Lett.}\ }\textbf {\bibinfo {volume}
  {113}},\ \bibinfo {pages} {171109} (\bibinfo {year} {2018})}\BibitemShut
  {NoStop}%
\bibitem [{\citenamefont {Trojek}\ and\ \citenamefont
  {Weinfurter}(2008)}]{trojek2008collinear}%
  \BibitemOpen
  \bibfield  {author} {\bibinfo {author} {\bibfnamefont {P.}~\bibnamefont
  {Trojek}}\ and\ \bibinfo {author} {\bibfnamefont {H.}~\bibnamefont
  {Weinfurter}},\ }\href {https://aip.scitation.org/doi/abs/10.1063/1.2924280}
  {\bibfield  {journal} {\bibinfo  {journal} {Appl. Phys. Lett.}\ }\textbf
  {\bibinfo {volume} {92}},\ \bibinfo {pages} {211103} (\bibinfo {year}
  {2008})}\BibitemShut {NoStop}%
\bibitem [{\citenamefont {Lohrmann}\ \emph {et~al.}(2019)\citenamefont
  {Lohrmann}, \citenamefont {Perumgatt},\ and\ \citenamefont
  {Ling}}]{lohrmann2019manipulation}%
  \BibitemOpen
  \bibfield  {author} {\bibinfo {author} {\bibfnamefont {A.}~\bibnamefont
  {Lohrmann}}, \bibinfo {author} {\bibfnamefont {C.}~\bibnamefont
  {Perumgatt}},\ and\ \bibinfo {author} {\bibfnamefont {A.}~\bibnamefont
  {Ling}},\ }\href
  {https://www.osapublishing.org/oe/abstract.cfm?uri=oe-27-10-13765} {\bibfield
   {journal} {\bibinfo  {journal} {Opt. Express}\ }\textbf {\bibinfo {volume}
  {27}},\ \bibinfo {pages} {13765–13772} (\bibinfo {year}
  {2019})}\BibitemShut {NoStop}%
\bibitem [{\citenamefont {Sachidananda}\ and\ \citenamefont
  {Ling}(2011)}]{sachidananda2011bias}%
  \BibitemOpen
  \bibfield  {author} {\bibinfo {author} {\bibfnamefont {S.}~\bibnamefont
  {Sachidananda}}\ and\ \bibinfo {author} {\bibfnamefont {A.}~\bibnamefont
  {Ling}},\ }\href {https://arxiv.org/pdf/1111.2656.pdf} {\bibfield  {journal}
  {\bibinfo  {journal} {arXiv preprint arXiv:1111.2656}\ } (\bibinfo {year}
  {2011})}\BibitemShut {NoStop}%
\bibitem [{\citenamefont {Clauser}\ \emph {et~al.}(1969)\citenamefont
  {Clauser}, \citenamefont {Horne}, \citenamefont {Shimony},\ and\
  \citenamefont {Holt}}]{clauser1969proposed}%
  \BibitemOpen
  \bibfield  {author} {\bibinfo {author} {\bibfnamefont {J.~F.}\ \bibnamefont
  {Clauser}}, \bibinfo {author} {\bibfnamefont {M.~A.}\ \bibnamefont {Horne}},
  \bibinfo {author} {\bibfnamefont {A.}~\bibnamefont {Shimony}},\ and\ \bibinfo
  {author} {\bibfnamefont {R.~A.}\ \bibnamefont {Holt}},\ }\href
  {https://journals.aps.org/prl/abstract/10.1103/PhysRevLett.23.880} {\bibfield
   {journal} {\bibinfo  {journal} {Phys. Rev. Lett.}\ }\textbf {\bibinfo
  {volume} {23}},\ \bibinfo {pages} {880–884} (\bibinfo {year}
  {1969})}\BibitemShut {NoStop}%
\bibitem [{\citenamefont {Lohrmann}\ \emph {et~al.}(2020)\citenamefont
  {Lohrmann}, \citenamefont {Perumangatt}, \citenamefont {Villar},\ and\
  \citenamefont {Ling}}]{lohrmanfeaturedpaper}%
  \BibitemOpen
  \bibfield  {author} {\bibinfo {author} {\bibfnamefont {A.}~\bibnamefont
  {Lohrmann}}, \bibinfo {author} {\bibfnamefont {C.}~\bibnamefont
  {Perumangatt}}, \bibinfo {author} {\bibfnamefont {A.}~\bibnamefont
  {Villar}},\ and\ \bibinfo {author} {\bibfnamefont {A.}~\bibnamefont {Ling}},\
  }\href {https://doi.org/10.1063/1.5124416} {\bibfield  {journal} {\bibinfo
  {journal} {Appl. Phys. Lett.}\ }\textbf {\bibinfo {volume} {116}},\ \bibinfo
  {pages} {021101} (\bibinfo {year} {2020})},\ \Eprint
  {https://arxiv.org/abs/https://doi.org/10.1063/1.5124416}
  {https://doi.org/10.1063/1.5124416} \BibitemShut {NoStop}%
\bibitem [{\citenamefont {Storm}\ \emph {et~al.}(2017)\citenamefont {Storm},
  \citenamefont {Cao}, \citenamefont {Litvinovitch}, \citenamefont
  {Puffenberger}, \citenamefont {Albert}, \citenamefont {Young}, \citenamefont
  {Pachowicz},\ and\ \citenamefont {Deely}}]{storm2017cubesat}%
  \BibitemOpen
  \bibfield  {author} {\bibinfo {author} {\bibfnamefont {M.}~\bibnamefont
  {Storm}}, \bibinfo {author} {\bibfnamefont {H.}~\bibnamefont {Cao}}, \bibinfo
  {author} {\bibfnamefont {S.}~\bibnamefont {Litvinovitch}}, \bibinfo {author}
  {\bibfnamefont {K.}~\bibnamefont {Puffenberger}}, \bibinfo {author}
  {\bibfnamefont {M.}~\bibnamefont {Albert}}, \bibinfo {author} {\bibfnamefont
  {J.}~\bibnamefont {Young}}, \bibinfo {author} {\bibfnamefont
  {D.}~\bibnamefont {Pachowicz}},\ and\ \bibinfo {author} {\bibfnamefont
  {T.}~\bibnamefont {Deely}},\ }in\ \href
  {https://digitalcommons.usu.edu/smallsat/2017/all2017/153/} {\emph {\bibinfo
  {booktitle} {Big Data From Small Satellites 2}}},\ \bibinfo {series and
  number} {\bibinfo {number} {SSC17-XI-09}}\ (\bibinfo {organization} {Proc. of
  the AIAA/USU Conf. on Small Satell.},\ \bibinfo {year} {2017})\ p.\ \bibinfo
  {pages} {1–4}\BibitemShut {NoStop}%
\bibitem [{\citenamefont {Jennewein}\ \emph {et~al.}(2014)\citenamefont
  {Jennewein}, \citenamefont {Grant}, \citenamefont {Choi}, \citenamefont
  {Pugh}, \citenamefont {Holloway}, \citenamefont {Bourgoin}, \citenamefont
  {Hakima}, \citenamefont {Higgins},\ and\ \citenamefont
  {Zee}}]{jennewein2014nanoqey}%
  \BibitemOpen
  \bibfield  {author} {\bibinfo {author} {\bibfnamefont {T.}~\bibnamefont
  {Jennewein}}, \bibinfo {author} {\bibfnamefont {C.}~\bibnamefont {Grant}},
  \bibinfo {author} {\bibfnamefont {E.}~\bibnamefont {Choi}}, \bibinfo {author}
  {\bibfnamefont {C.}~\bibnamefont {Pugh}}, \bibinfo {author} {\bibfnamefont
  {C.}~\bibnamefont {Holloway}}, \bibinfo {author} {\bibfnamefont {J.-P.}\
  \bibnamefont {Bourgoin}}, \bibinfo {author} {\bibfnamefont {H.}~\bibnamefont
  {Hakima}}, \bibinfo {author} {\bibfnamefont {B.}~\bibnamefont {Higgins}},\
  and\ \bibinfo {author} {\bibfnamefont {R.}~\bibnamefont {Zee}},\ }in\ \href
  {https://doi.org/10.1117/12.2067548} {\emph {\bibinfo {booktitle} {Emerging
  Technologies in Security and Defence II; and Quantum-Physics-based
  Information Security III}}},\ \bibinfo {series and number} {\bibinfo {number}
  {925401}}\ (\bibinfo {organization} {International Society for Optics and
  Photonics},\ \bibinfo {year} {2014})\ p.\ \bibinfo {pages}
  {1–6}\BibitemShut {NoStop}%
\bibitem [{\citenamefont {Kerstel}\ \emph {et~al.}(2018)\citenamefont
  {Kerstel}, \citenamefont {Gardelein}, \citenamefont {Barthelemy},
  \citenamefont {{The CSUG Team}}, \citenamefont {Fink}, \citenamefont
  {Joshi},\ and\ \citenamefont {Ursin}}]{kerstel2018nanobob}%
  \BibitemOpen
  \bibfield  {author} {\bibinfo {author} {\bibfnamefont {E.}~\bibnamefont
  {Kerstel}}, \bibinfo {author} {\bibfnamefont {A.}~\bibnamefont {Gardelein}},
  \bibinfo {author} {\bibfnamefont {M.}~\bibnamefont {Barthelemy}}, \bibinfo
  {author} {\bibnamefont {{The CSUG Team}}}, \bibinfo {author} {\bibfnamefont
  {M.}~\bibnamefont {Fink}}, \bibinfo {author} {\bibfnamefont {S.~K.}\
  \bibnamefont {Joshi}},\ and\ \bibinfo {author} {\bibfnamefont
  {R.}~\bibnamefont {Ursin}},\ }\href
  {https://link.springer.com/article/10.1140/epjqt/s40507-018-0070-7}
  {\bibfield  {journal} {\bibinfo  {journal} {EPJ Quantum Technol.}\ }\textbf
  {\bibinfo {volume} {5}},\ \bibinfo {pages} {6} (\bibinfo {year}
  {2018})}\BibitemShut {NoStop}%
\bibitem [{\citenamefont {Haber}\ \emph {et~al.}(2018)\citenamefont {Haber},
  \citenamefont {Garbe}, \citenamefont {Schilling},\ and\ \citenamefont
  {Rosenfeld}}]{haber2018qube}%
  \BibitemOpen
  \bibfield  {author} {\bibinfo {author} {\bibfnamefont {R.}~\bibnamefont
  {Haber}}, \bibinfo {author} {\bibfnamefont {D.}~\bibnamefont {Garbe}},
  \bibinfo {author} {\bibfnamefont {K.}~\bibnamefont {Schilling}},\ and\
  \bibinfo {author} {\bibfnamefont {W.}~\bibnamefont {Rosenfeld}},\ }in\ \href
  {https://digitalcommons.usu.edu/cgi/viewcontent.cgi?article=4081&=&context=smallsat&=&sei-redir=1&referer=https%253A%252F%252Fscholar.google.com.sg%252Fscholar%253Fhl%253Den%2526as_sdt%253D0%25252C5%2526q%253DQUBE%2525E2%252580%252594A%252BCubeSat%252Bfor%252BQuantum%252BKey%252BDistribution%252BExperiments%2526btnG%253D#search=%22QUBE—A%20CubeSat%20Quantum%20Key%20Distribution%20Experiments%22}
  {\emph {\bibinfo {booktitle} {Advanced Technologies I}}},\ \bibinfo {series
  and number} {\bibinfo {number} {SSC18-III-05}}\ (\bibinfo {organization}
  {Proc. of the AIAA/USU Conf. on Small Satell.},\ \bibinfo {year} {2018})\ p.\
  \bibinfo {pages} {1–8}\BibitemShut {NoStop}%
\bibitem [{\citenamefont {Lee}\ \emph {et~al.}(2019)\citenamefont {Lee},
  \citenamefont {Shen}, \citenamefont {Cerè}, \citenamefont {Troupe},
  \citenamefont {Lamas-Linares},\ and\ \citenamefont {Kurtsiefer}}]{lee19}%
  \BibitemOpen
  \bibfield  {author} {\bibinfo {author} {\bibfnamefont {J.}~\bibnamefont
  {Lee}}, \bibinfo {author} {\bibfnamefont {L.}~\bibnamefont {Shen}}, \bibinfo
  {author} {\bibfnamefont {A.}~\bibnamefont {Cerè}}, \bibinfo {author}
  {\bibfnamefont {J.}~\bibnamefont {Troupe}}, \bibinfo {author} {\bibfnamefont
  {A.}~\bibnamefont {Lamas-Linares}},\ and\ \bibinfo {author} {\bibfnamefont
  {C.}~\bibnamefont {Kurtsiefer}},\ }\href {https://doi.org/10.1063/1.5086493}
  {\bibfield  {journal} {\bibinfo  {journal} {Appl. Phys. Lett.}\ }\textbf
  {\bibinfo {volume} {114}},\ \bibinfo {pages} {101102} (\bibinfo {year}
  {2019})}\BibitemShut {NoStop}%
\bibitem [{\citenamefont {Naughton}\ \emph {et~al.}(2019)\citenamefont
  {Naughton}, \citenamefont {Bedington}, \citenamefont {Barraclough},
  \citenamefont {Islam}, \citenamefont {Griffin}, \citenamefont {Smith},
  \citenamefont {Kurtz}, \citenamefont {Alenin}, \citenamefont {Vaughn},
  \citenamefont {Ramana}, \citenamefont {Dimitrijevic}, \citenamefont {Tang},
  \citenamefont {Kurtsiefer}, \citenamefont {Ling},\ and\ \citenamefont
  {Boyce}}]{naughton2019design}%
  \BibitemOpen
  \bibfield  {author} {\bibinfo {author} {\bibfnamefont {D.~P.}\ \bibnamefont
  {Naughton}}, \bibinfo {author} {\bibfnamefont {R.}~\bibnamefont {Bedington}},
  \bibinfo {author} {\bibfnamefont {S.}~\bibnamefont {Barraclough}}, \bibinfo
  {author} {\bibfnamefont {T.}~\bibnamefont {Islam}}, \bibinfo {author}
  {\bibfnamefont {D.}~\bibnamefont {Griffin}}, \bibinfo {author} {\bibfnamefont
  {B.}~\bibnamefont {Smith}}, \bibinfo {author} {\bibfnamefont
  {J.}~\bibnamefont {Kurtz}}, \bibinfo {author} {\bibfnamefont {A.~S.}\
  \bibnamefont {Alenin}}, \bibinfo {author} {\bibfnamefont {I.~J.}\
  \bibnamefont {Vaughn}}, \bibinfo {author} {\bibfnamefont {A.}~\bibnamefont
  {Ramana}}, \bibinfo {author} {\bibfnamefont {I.}~\bibnamefont
  {Dimitrijevic}}, \bibinfo {author} {\bibfnamefont {Z.~S.}\ \bibnamefont
  {Tang}}, \bibinfo {author} {\bibfnamefont {C.}~\bibnamefont {Kurtsiefer}},
  \bibinfo {author} {\bibfnamefont {A.}~\bibnamefont {Ling}},\ and\ \bibinfo
  {author} {\bibfnamefont {R.}~\bibnamefont {Boyce}},\ }\href
  {https://www.spiedigitallibrary.org/journals/Optical-Engineering/volume-58/issue-1/016106/Design-considerations-for-an-optical-link-supporting-intersatellite-quantum-key/10.1117/1.OE.58.1.016106.short}
  {\bibfield  {journal} {\bibinfo  {journal} {Opt. Eng.}\ }\textbf {\bibinfo
  {volume} {58}},\ \bibinfo {pages} {016106} (\bibinfo {year}
  {2019})}\BibitemShut {NoStop}%
\bibitem [{\citenamefont {Tang}(2018)}]{tang2018towards}%
  \BibitemOpen
  \bibfield  {author} {\bibinfo {author} {\bibfnamefont {Z.~X.}\ \bibnamefont
  {Tang}},\ }\emph {\bibinfo {title} {Towards a Space-Based Quantum Key
  Distribution Network: Developing a Miniaturised Polarisation Entangled Photon
  Pair Source for Nanosatellites}},\ \href
  {https://scholarbank.nus.edu.sg/handle/10635/146939} {Ph.D. thesis},\
  \bibinfo  {school} {National University of Singapore (Singapore)} (\bibinfo
  {year} {2018})\BibitemShut {NoStop}%
\bibitem [{\citenamefont {Wang}\ \emph {et~al.}(2014)\citenamefont {Wang},
  \citenamefont {Li}, \citenamefont {Cui}, \citenamefont {Wang}, \citenamefont
  {Sun},\ and\ \citenamefont {Cheng}}]{wang2014operations}%
  \BibitemOpen
  \bibfield  {author} {\bibinfo {author} {\bibfnamefont {K.}~\bibnamefont
  {Wang}}, \bibinfo {author} {\bibfnamefont {J.}~\bibnamefont {Li}}, \bibinfo
  {author} {\bibfnamefont {Z.}~\bibnamefont {Cui}}, \bibinfo {author}
  {\bibfnamefont {N.}~\bibnamefont {Wang}}, \bibinfo {author} {\bibfnamefont
  {Q.}~\bibnamefont {Sun}},\ and\ \bibinfo {author} {\bibfnamefont
  {L.}~\bibnamefont {Cheng}},\ }\href
  {https://www.sciencedirect.com/science/article/pii/S0168900214009309}
  {\bibfield  {journal} {\bibinfo  {journal} {Nucl. Instrum. Methods Phys.
  Res.}\ }\textbf {\bibinfo {volume} {767}},\ \bibinfo {pages} {235–244}
  (\bibinfo {year} {2014})}\BibitemShut {NoStop}%
\bibitem [{\citenamefont {Villar}(2019)}]{zafra2019building}%
  \BibitemOpen
  \bibfield  {author} {\bibinfo {author} {\bibfnamefont {A.}~\bibnamefont
  {Villar}},\ }\emph {\bibinfo {title} {Building entangled photon pair sources
  for quantum key distribution with nano-satellites}},\ \href
  {https://sb-cris.nus.edu.sg/handle/10635/161246} {Ph.D. thesis},\ \bibinfo
  {school} {National University of Singapore (Singapore)} (\bibinfo {year}
  {2019})\BibitemShut {NoStop}%
\end{thebibliography}%

\clearpage
\onecolumngrid
\begin{appendices}
\renewcommand\thefigure{\thesection.\arabic{figure}}
\renewcommand{\theHfigure}{A\arabic{figure}}
\setcounter{figure}{0}

\section{Angular tolerance of the SPDC crystals}
\label{appendix:tolerance}

\begin{figure}[h]
	\centering
	\includegraphics{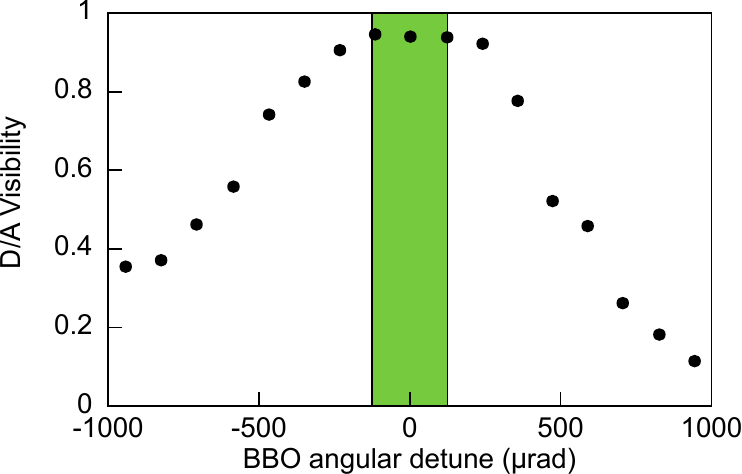}
	\caption{Visibility in the diagonal/anti-diagonal ($D/A$) basis with respect to angular detuning of a BBO crystal. The green area depicts the design angular tolerance ($\pm$\SI{100}{\micro\radian}). This requirement is met by the optical bench~\cite{tang2018towards}.}
	\label{fig:SI-1}
\end{figure}

The phase-matching conditions for the BBO crystals used in the source are achieved by angle tuning. With this technique, the relative angle between the pump beam and the optical axis of the crystal is crucial. This angle must be maintained within $\pm$\SI{100}{\micro \radian} to maintain source performance (e.g. photon-pair phase). This is depicted in Fig.~\ref{fig:SI-1}, where the visibility in the $D/A$ basis is measured while introducing an angular detuning in one of the two BBO crystals.

\section{In-orbit heater operation}
\label{appendix:temps}

\begin{figure}[h]
	\centering
	\includegraphics{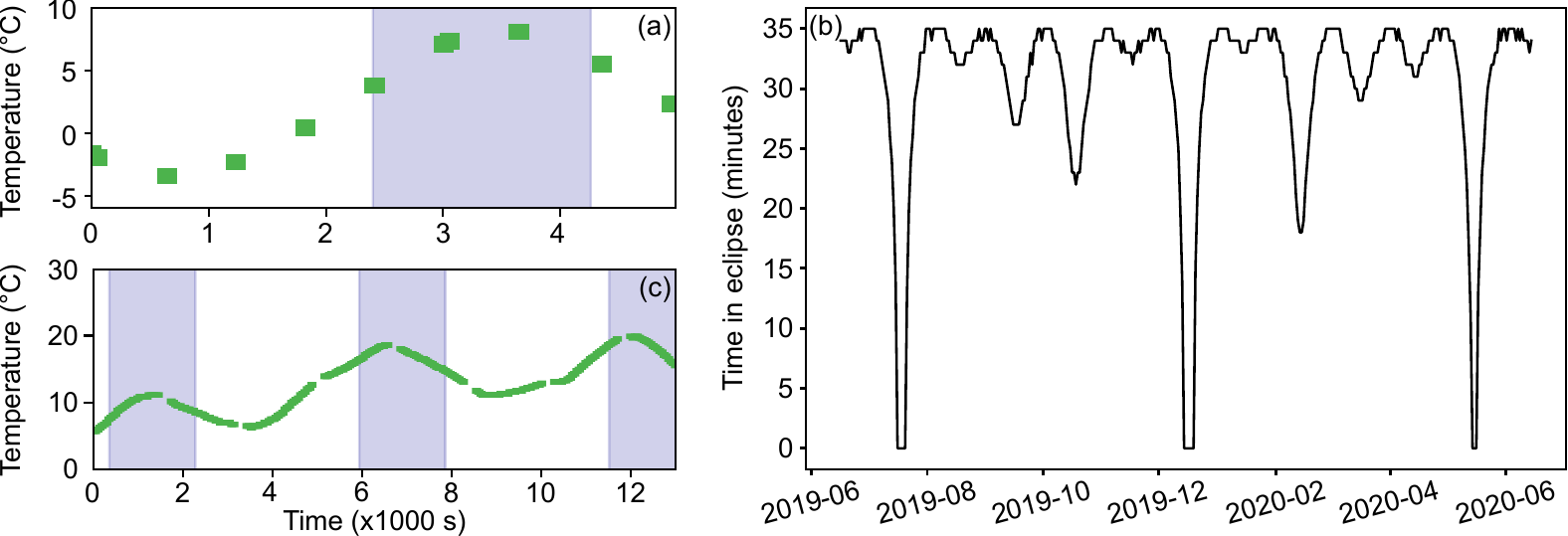}
	\caption{(a) Temperature of the experiment during orbit. The shaded area illustrates the time spent in eclipse. (b) Solar illumination conditions of SpooQy-1 during its expected lifetime. (c) Temperature of the experiment when the heater is operated. There is a 120 second gap in between heating cycles for the on-board electronics to perform system checks.}
	\label{fig:SI-2}
\end{figure}

The safe operating temperature range for the experiment was defined as
between \SI{15}{\degree C} to \SI{28}{\degree C}; this was driven by the requirements of the pump laser diode. Most of the time, the experimental apparatus does not experience
this temperature. Instead, the temperature fluctuates between \SI{-5}{\degree C} and
\SI{10}{\degree C} (Fig.~\ref{fig:SI-2}(a)). These fluctuations depend on the satellite's position and orientation during orbit. Furthermore, during the lifetime of the
satellite, the solar illumination condition varies as depicted in Fig.~\ref{fig:SI-2}(b). Due to the specific inclination of the orbit~\cite{wang2014operations}, in some cases
the satellite does not spend time in eclipse for several days (note the
pronounced valleys in Fig.~\ref{fig:SI-2}(b)). During these non-eclipse periods, the
satellite could heat up, potentially damaging the experimental apparatus
with excess heat.

In order to restrict heat conduction between the experimental apparatus
and the satellite bus, and also maintain passive optical stability
across varying temperatures, an isostatic mount was fabricated. This
mount is made out of three \SI{0.4}{mm} thick stainless steel blades
(manuscript under preparation). The blades serve to absorb any thermal
expansion mismatch between the experiment and the satellite structure.
To achieve the necessary operating temperature a \SI{2.5}{\watt} heater is
activated until the required condition is achieved (see Fig.~\ref{fig:SI-2}(c)).

\section{Survey of pump modes that produce high-quality entanglement}
\label{appendix:heatmap}

\begin{figure}[h]
	\centering
	\includegraphics{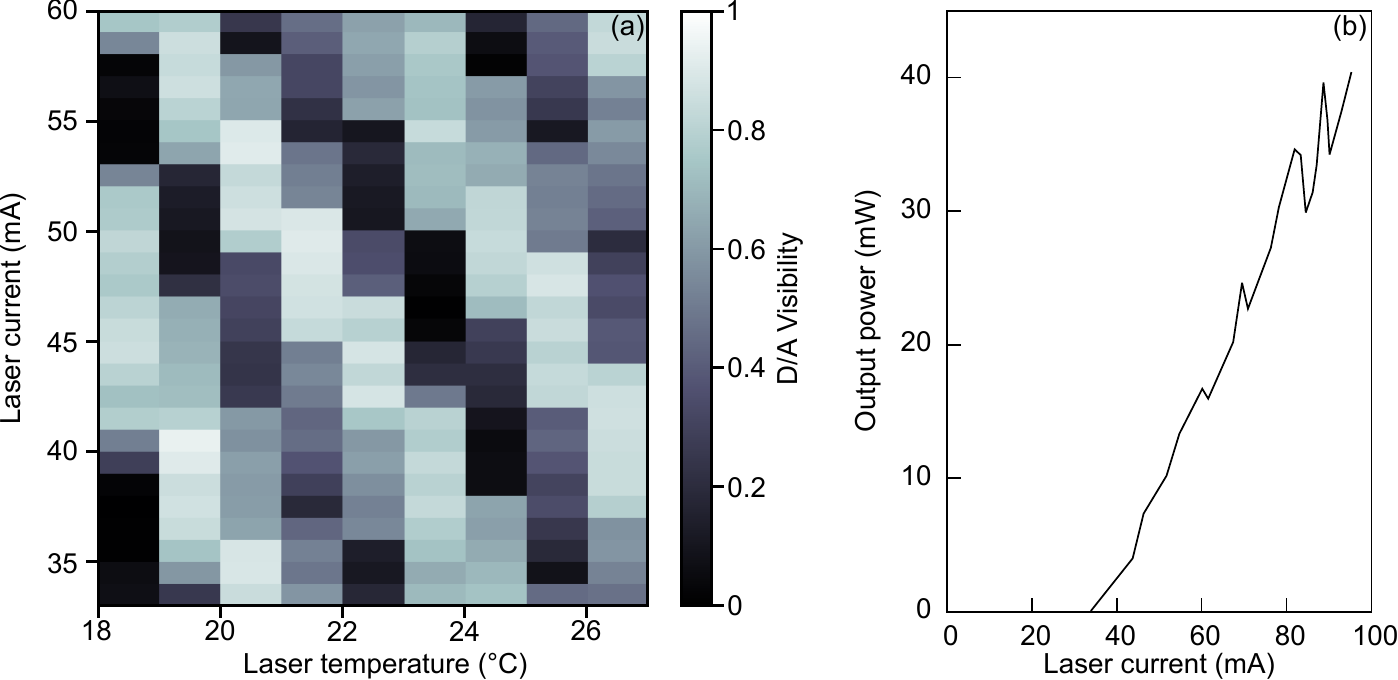}
	\caption{(a) Visibility in the $D/A$ basis for different combinations of laser currents and temperatures. (b) Laser current versus output power of the pump laser utilised in the experiment.}
	\label{fig:SI-3}
\end{figure}

The pump wavelength is a function of temperature and current~\cite{zafra2019building}. Changes in wavelength (\textit{mode-hops}) are accompanied by changes in phase $\Delta\varphi$. This phase change can degrade the entanglement quality produced by the source. Mode-hops are common in orbit due to the fluctuating temperature. To recover the entanglement quality, an optimal laser current can be used. A survey of the in-orbit pump laser was performed at different temperatures to identify laser currents that supported the production of high-quality entanglement. The resulting heatmap (see Fig.~\ref{fig:SI-3}(a)) was used as a reference when operating the experiment in space. Additionally, it is worth noting that the output power of the laser diode does not always scale proportionally with the laser current, as shown in Fig.~\ref{fig:SI-3}(b).

\newpage
\section{Satellite Bus and Ground Control}
\label{appendix:gs}
The SpooQy-1 satellite (with NORAD catalogue number: 44332) is based on a 3U Gom-X platform from GomSpace ApS.
The SpooQy-1 satellite bus includes: a half-duplex UHF transceiver combined with deployable canted turnstile UHF antennas used for both uplink and downlink; a 32-bit AVR computer with a \SI{64}{\mega Byte} flash storage used as the onboard computer (OBC) for housekeeping and data handling; an attitude determination system (embedded in the OBC) with 3 magnetorquers for 3-axis detumbling; and a \SI{38}{\watt hr} battery pack (four lithium-ion 18650 cells, \SI{7.7}{\watt hr} maximum depth of discharge) with the electrical power management system. 

The peak system power consumption is rated at \SI{3.85}{\watt}, while the peak power consumption of the experiment is rated at \SI{2.5}{\watt}. As photon detection is performed on board the satellite, no optical ground station is needed and only UHF ground stations are used for telemetry and satellite command. To increase the link budget two ground stations were used; one located at the National University of Singapore (NUS) campus in Singapore, and another one at the University of Applied Sciences Northwestern Switzerland (FHNW) campus in Switzerland. Both ground stations are equipped with two WiMo X-Quad antennas (amplification gain of \SI{15}{\decibel i}). The rotor for the tracking mount is controlled by a Linux-based server computer (NanoCom MS100). The ground station radio (NanoCom GS100) is the ground counterpart (with a \SI{25}{\watt} power amplifier) to the NanoCom AX100 radio on board SpooQy-1, designed to work together using the CubeSat Space Protocol.

\clearpage
\section{Geometrical loss}
\label{appendix:mode-mismatch}

\begin{figure*}[h]
	\centering
	\includegraphics{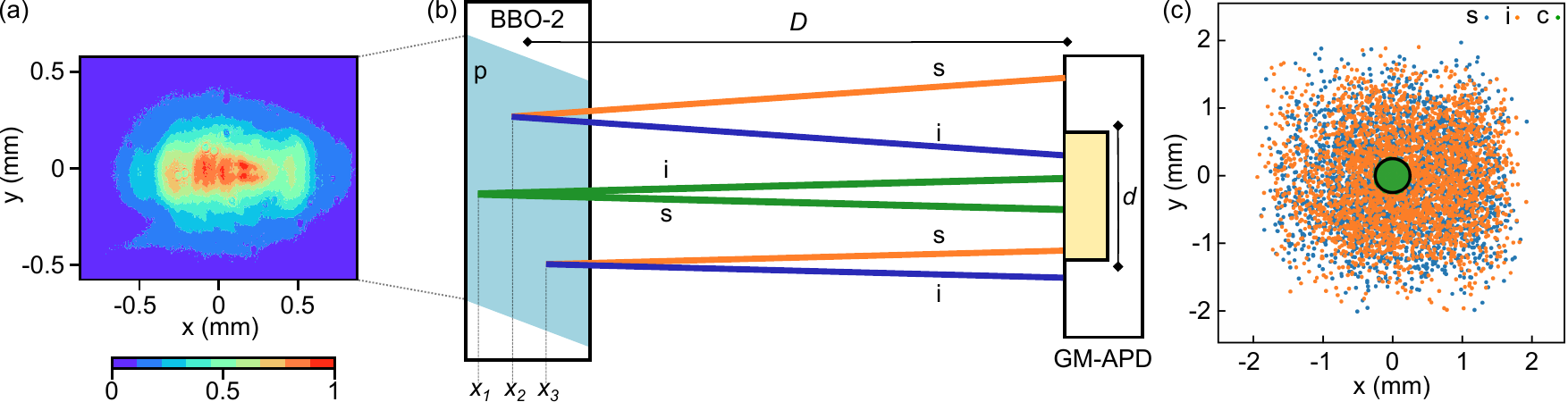}
	\caption{(a) Intensity profile of the laser beam cross-section recorded with a charge-coupled device camera. The collimated laser beam has a full-width half-maximum of \SI{800}{\micro\metre}$\times$\SI{400}{\micro\metre}. (b) Sketch of the geometrical loss. The generated entangled photon-pairs are not always successfully detected due to their intrinsic emission angle and the different generation positions within the crystal ($x_i$). For a pair to be successfully detected (and to generate a coincidence count), both signal and idler photons need to hit the active area of the detectors (e.g. green rays). If only one of the photons is detected (e.g. orange or blue rays), only a single count will be recorded. For clarity, only one detector and no wavelength separation of the photons have been sketched. (c) A ray-tracing simulation to calculate geometrical loss. The black circle depicts the \SI{500}{\micro\metre} diameter of the detector active area. Entangled photon-pairs hitting that region will be registered as coincidences. This simulation predicts a geometric efficiency of $\leq$4\%. p: pump; s: signal; i: idler; c: coincidence; d: \SI{500}{\micro\metre}; D: \SI{100}{\milli\metre}; GM-APD: Geiger-mode avalanche photodiode.}
	\label{fig:SI-4}
\end{figure*}

The decision to forego the use of collection lenses in the experiment leads to an optical configuration in which the geometrical loss dominates the overall performance of the experiment. This is illustrated in Fig.~\ref{fig:SI-4}. As the laser beam (Fig.~\ref{fig:SI-4}(a)) triggers SPDC along the nonlinear crystals, entangled photon-pairs are emitted and directed towards the Geiger-mode avalanche photodiodes (GM-APD). Depending on both the position within the crystal at which the photon-pair is generated and its emission opening angle (up to $0.3 \SI{}{\degree}$), the percentage of photon-pairs successfully detected can be estimated via ray tracing techniques.

We use the intensity distribution of the pump beam in one crystal (BBO-2) to randomly generate rays of downconverted photon-pairs. Signal and idler wavelengths and opening angles are chosen randomly distributed based on phase-matching considerations. We propagate the individual rays (considering the refraction at the crystal-air interface) towards the single photon detectors (Fig.~\ref{fig:SI-4}(b)). The rays are discarded if none of the photons hits its corresponding detector. The ray tracing results are shown in Fig.~\ref{fig:SI-4}(c). It can be seen how only a small percentage ($\leq$4\%; green, central region of the plot) of the generated pairs is successfully detected. Here, a successful photon-pair detection includes both signal and idler photons reaching the active area of the GM-APDs.

Geometrical loss can be mitigated with appropriate collection lenses and state-of-the-art coincidence rates using the identical source configuration can be achieved~\cite{lohrmann2018high}.

\end{appendices}

\end{document}